\documentstyle[12pt]{article}
\begin{document}

\begin{titlepage}

\vspace{-4cm}

\title{
   {\LARGE The $B$-$L$/Electroweak Hierarchy \\[.1in]
    in Heterotic String and M-Theory\\[.5cm]  }}
                       
\author{{\bf
   Michael Ambroso$^{2}$ 
   and Burt A.~Ovrut$^{1,2}$}\\[5mm]
   {\it $^1$School of Natural Sciences, The Institute for Advanced Study} \\
   {\it Princeton, New Jersey, 08540} \\[3mm]
   {\it $^2$Department of Physics, University of Pennsylvania} \\
   {\it Philadelphia, PA 19104--6396}}
\date{}

\maketitle

\begin{abstract}
\noindent
$E_{8} \times E_{8}$ heterotic string and M-theory, when compactified on a Calabi-Yau threefold admitting an $SU(4)$ vector bundle with Wilson lines, can give rise to the exact MSSM spectrum with three right-handed neutrino chiral superields, one per family. Rank preserving Wilson lines require that the standard model group be augmented by a gauged $U(1)_{B-L}$. Since there are no fields in this theory for which $3(B-L)$ is an even, non-zero integer, the gauged $B$-$L$ symmetry must be spontaneously broken at a low scale, not too far above the electroweak scale. It is shown that in these heterotic standard models, the $B$-$L$ symmetry can be broken, with a phenomenologically 
viable $B$-$L$/electroweak hierarchy, by at least one right-handed sneutrino acquiring a vacuum expectation value. This is explicitly demonstrated, in a specific region of parameter space, using a renormalization group analysis and soft supersymmetry breaking operators. The vacuum state is shown to be a stable, local minimum of the potential and the resultant hierarchy is
explicitly presented in terms of $tan\beta$. 

\vspace{.3in}
\noindent
\end{abstract}

\thispagestyle{empty}

\end{titlepage}

An important goal of heterotic superstrings and $M$-theory is to show that these higher-dimensional theories can be ``compactified'' to four-dimensional, phenomenologically realistic particle physics. Specifically, one would like to prove that the minimal ${\cal{N}}=1$ supersymmetric standard model (MSSM), modified by the addition of three right-handed neutrino chiral supermultiplets, one per family, can arise in this manner. The necessity of having three right-handed neutrino supermultiplets puts strong constraints on heterotic model building. A natural way to achieve this is to compactify on smooth Calabi-Yau threefolds that admit slope-stable, holomorphic vector bundles with structure group $SU(4)$. The non-vanishing connections associated with these bundles then spontaneously break the $E_{8}$ group of the heterotic theory down to $Spin(10)$. Each $\bf 16$ representation of $Spin(10)$ contains a complete family of quarks/leptons plus a right-handed neutrino, exactly as required. 

A second requirement is that the four-dimensional theory be symmetric, at least to a low energy scale, under $R$-parity \cite{1a,2a} or, equivalently in a supersymmetric theory, matter-parity. This ${\bf Z}_{2}$ symmetry prohibits dangerous baryon and lepton violating processes, such as rapid nucleon decay.
The requirement of $R$-parity, however, also puts additional strong constraints on heterotic model building. While it is difficult in realistic smooth heterotic compactifications to obtain a ${\bf Z}_{2}$ symmetry of the four-dimensional theory, in particular of  the soft supersymmetry breaking interactions, it is straightforward to extend the standard model group by a gauged $U(1)_{B-L}$, which contains matter-parity. Models of this type have been proposed within the context of field theory \cite{1b,1,2,3,4}  and some string theories, such as heterotic orbifolds \cite{raby}, in which, in addition to the MSSM matter spectrum, new chiral fields are added for which $3(B-L)$ is an even, non-zero integer. These new fields can acquire vacuum expectation values (VEVs) which, while spontaneously breaking gauged $B$-$L$ symmetry at a high scale, preserve the matter-parity subgroup. Unfortunately, this is {\it never possible} in realistic smooth compactifications of heterotic theory, since the $E_{8}$ decomposition under the vector bundle structure group never has representations satisfying this condition. It follows that, in {\it smooth} heterotic compactifications, one is forced to consider the remaining possibility; that is, that $U(1)_{B-L}$ is spontaneously broken by $3(B-L)$ odd fields at a scale not too far above the electroweak scale.  This will play the same role of suppressing  baryon and lepton number violating operators. 

In fact, a gauged $U(1)_{B-L}$ group arises naturally in the $Spin(10)$ models discussed above. The traditional way to break the rank five $Spin(10)$ to the standard model gauge group is by extending the $SU(4)$ bundle with Abelian Wilson lines. These, however, preserve the rank of the gauge group and, hence, the rank four standard model group must be extended by a product with a rank one group, precisely, it turns out, $U(1)_{B-L}$.
A class of smooth heterotic compactifications of this type were constructed in \cite{1A,2A}. Specifically, they compactify heterotic theory on elliptically fibered Calabi-Yau threefolds that admit a fixed-point free ${\bf Z_{3}} \times {\bf Z_{3}}$ isometry \cite{1AA,2AA}. Slope-stable holomorphic vector bundles with structure group $SU(4)$ were constructed over them which spontaneously break $E_{8}$ to $Spin(10)$. The Abelian 
${\bf Z_{3}} \times {\bf Z_{3}}$ Wilson lines then further break $Spin(10)$ to the low-energy gauge group
\begin{equation}
G=SU(3)_{C} \times SU(2)_{L} \times U(1)_{Y} \times U(1)_{B-L} \ .
\label{1}
\end{equation}
The manifold and vector bundles being Calabi-Yau and slope-stable, holomorphic respectively, assure that the four-dimensional theory is ${\cal{N}}=1$ supersymmetric. In addition to the vector superfields corresponding to the gauge group in (\ref{1}), the low-energy matter spectrum was found, using cohomology techniques introduced in \cite{1B,2B,3B},  to be three families of quark and lepton chiral superfields, each family with a right-handed neutrino. They transform under the gauge group in the standard manner as
\begin{eqnarray}
& &Q_{i}=({\bf 3},{\bf 2},1/3,1/3), \quad u_{i} =({\bf \bar{3}}, {\bf 1}, -4/3, -1/3), \quad d_{i}=({\bf \bar{3}}, {\bf 1}, 2/3, -1/3) \nonumber \\
& & \qquad \ \ \ L_{i}=({\bf 1},{\bf 2},-1,-1), \quad \nu_{i}=({\bf 1}, {\bf 1}, 0, 1), \quad e_{i}=({\bf 1}, {\bf 1}, 2, 1)
\label{2}
\end{eqnarray}
for the left  and right-handed squarks and leptons respectively, where $i=1,2,3$. The spectrum also contains one pair of Higgs-Higgs conjugate chiral superfields transforming as
\begin{equation}
H=({\bf 1},{\bf 2},1,0), \qquad \bar{H}=({\bf 1},{\bf 2}, -1,0).
\label{4}
\end{equation}
This is {\it precisely} the matter and Higgs spectrum of the MSSM.
In addition, the theory contains three Kahler moduli, three complex structure moduli and thirteen vector bundle moduli, all of which are uncharged under the gauge group (\ref{1}). 

The supersymmetric potential energy is given by the usual sum over the modulus squared of the $F$
and $D$-terms. The $F$-terms are determined from the most general superpotential  invariant under the gauge group,
\begin{equation}
W=\mu H\bar{H} +{\sum_{i,j=1}^{3}}\left(\lambda_{u, ij} Q_{i}Hu_{j}+\lambda_{d, ij} Q_{i}\bar{H}d_{j}+\lambda_{\nu, ij} L_{i}H\nu_{j}+\lambda_{e, ij} L_{i}\bar{H}e_{j}\right)
\label{5}
\end{equation}
Note that the dangerous lepton and baryon number violating interactions
\begin{equation}
L_{i}L_{j}e_{k}, \quad L_{i}Q_{j}d_{k}, \quad u_{i}d_{j}d_{k} 
\label{6}
\end{equation}
which generically would lead to rapid nucleon decay, are disallowed by the $U(1)_{B-L}$ gauge symmetry. To simplify the calculations, we will assume a mass-diagonal basis where $\lambda_{u, ij}=\lambda_{d, ij}=\lambda_{\nu, ij}=\lambda_{e, ij}=0$ for $i \neq j$
and denote the diagonal Yukawa couplings by $\lambda_{ii}=\lambda_{i}$, $i=1,2,3$. A constant, field-independent $\mu$ parameter cannot arise in a supersymmetric string vacuum  since the Higgs fields are zero modes. However, the $H{\bar{H}}$ bilinear can have higher-dimensional couplings to moduli through both holomorphic and non-holomorphic interactions in the superpotential and Kahler potential respectively. When moduli acquire VEVs due to non-perturbative effects, these can induce non-vanishing supersymmetric contributions to $\mu$. A non-zero $\mu$ can also be generated by gaugino condensation in the hidden sector. Why this induced $\mu$-term should be small enough to be consistent with electroweak symmetry breaking is a difficult, model dependent problem. In this paper, we will not discuss this ``$\mu$-problem'', but simply assume that the $\mu$ parameter is at, or below, the electroweak scale. In fact, so as {\it to empasize the $B$-$L$/electroweak hierarchy and simplify the calculation, we will take $\mu$, while non-zero, to be substantially smaller than the electroweak scale}, making its effect  sub-dominant. This can be implemented consistently throughout the entire scaling regime.

The $SU(3)_{C}$ and $SU(2)_{L}$ $D$-terms are of the standard form, while
\begin{eqnarray}
D_{Y} & = & \xi_{Y} +g_{Y}{{\phi}_{A}}^{\dagger}\left({\bf{Y}\rm}/2\right)_{AB}{\phi}_{B} \ , \label{7} \\
D_{B-L} &= & \xi_{B-L} +g_{B-L}{\phi}_{A}^{\dagger}\left({\bf{Y_{B-L}}\rm}\right)_{AB}{\phi}_{B} 
\nonumber
\end{eqnarray}
where the index $A$ runs over all scalar fields ${\phi}_{A}$. 
Note that each of these Abelian $D$-terms potentially has a Fayet-Iliopoulos (FI) additive constant.
As with the $\mu$ parameter, constant field-independent FI terms cannot occur in string vacua since the low energy fields are zero modes. Field-dependent FI terms can occur in some contexts, see for example \cite{Lara}. However, since both the hypercharge and $B$-$L$ gauge symmetries are anomaly free, such field-dependent FI  terms are not generated in the supersymmetric effective theory. We include them in (\ref{7}) since they can, in principle, arise at a lower scale from radiative corrections once supersymmetry is softly broken \cite{Jones}. Be that as it may, if calculations are done {\it in the $D$-eliminated formalism, which we use in this paper, these FI parameters can be consistently absorbed into the definition of the soft scalar masses} and their beta functions. Hence, we will no longer consider them.

In addition to the supersymmetric potential, the Lagrangian density also contains explicit ``soft'' supersymmetry violating terms \cite{1C}. Those relevant to this paper are $V_{\rm soft}=V_{2s}+V_{2f}$,
where $V_{2s}$ are the scalar quadratic terms 
\begin{eqnarray}
V_{2s} & = & { \sum_{i=1}^{3}} (m^{2}_{Q_{i}}|{Q}_{i}|^{2}+m^{2}_{u_{i}}|{u}_{i}|^{2}+
                        m^{2}_{d_{i}}|{d}_{i}|^{2} 
                  + m^{2}_{L_{i}}|{L}_{i}|^{2}+m^{2}_{\nu_{i}}|{\nu}_{i}|^{2}
                        \label{9}  \nonumber \\   
          & & +m^{2}_{e_{i}}|{e}_{i}|^{2})+m_{H}^{2}|H|^{2} +m_{\bar{H}}^{2}|\bar{H}|^{2}  -(BH\bar{H}+hc), \label{8}  
\end{eqnarray}
and $V_{2f}$ contains the gaugino mass terms
\begin{equation}
V_{2f}= \frac{1}{2} M_{3} \lambda_{3} \lambda_{3}+ \dots 
hc.
\label{10}
\end{equation}
As above, we have taken the parameters in (\ref{9}) and (\ref{10}) to be flavor-diagonal. Cubic scalar interactions as well as the remaining gaugino mass terms can be chosen small enough to be ignored in this calculation, as discussed below.

The heterotic compactifications described here satisfy the two criteria discussed above; that is, they give 
softly broken ${\cal{N}}=1$ supersymmetric theories with exactly the MSSM matter spectrum with three right-handed neutrinos, and their gauge group extends the standard model group by precisely a factor of $U(1)_{B-L}$.   
However, to be realistic, these theories must spontaneously break the $U(1)_{B-L}$ symmetry not too far above the electroweak scale. Clearly, this can only be accomplished if at least one of the right-handed sneutrinos develops a non-vanishing vacuum expectation value. It is straightforward to show using (\ref{5}), (\ref{7}) and (\ref{8}) that, assuming one is free to choose all parameters at the electroweak scale, both $U(1)_{B-L}$ and electroweak symmetry can be broken with a realistic hierarchy between them. Quintessentially, however, one is {\it not} free to so choose the parameters. As is well-known, their initial values just below the compactification scale are set by the geometric and bundle moduli expectation values \cite{1c,2c,3c,3cc,4c}. At any lower scale, the parameters are determined by a complicated set of intertwined, non-linear renormalization group equations (RGEs) \cite{1D,2D,V,3D,5D,6D,7D}. Even if one chooses the initial values arbitrarily, it is unclear that these will allow for an appropriate spontaneous breakdown of both the $U(1)_{B-L}$ and electroweak symmetries. There are many potential problems that can occur. These range, for example, from inducing color or charge breaking expectation values, to not being able to break $B$-$L$ at all, to breaking $B$-$L$ but inducing a correlation with electroweak breaking that is unphysical, such as the electroweak scale being much larger than the $B$-$L$ scale, and so on. All of these scenarios are easily realized. To prove that both $U(1)_{B-L}$ and electroweak symmetries can be broken with an appropriate hierarchy, one must show this explicitly by solving the RGEs for a specific choice of initial parameters. In this paper, we present the results of a quasi-analytic solution of the renormalization group equations valid for a restricted range of parameter space. The detailed calculations will be given elsewhere \cite{1E}. It will be shown in \cite{1E} that initial parameters can be chosen so that $U(1)_{Y}$ and $U(1)_{B-L}$ kinetic mixing \cite{1e,2e} is small. Hence, we ignore such mixing in this paper. This solution demonstrates that {\it an appropriate $B$-$L$/electroweak hierarchy can indeed be achieved for a range of initial parameters}. We have backed up these results with explicit numerical solutions of the RGEs that will be presented elsewhere.

We begin our analysis with the renormalization group solution for the gauge parameters, $g_{a}$, $a=1,\dots,4$, chosen so as to unify to $g(0) \simeq .726$ at the scale $M_{u}\simeq 3 \times10^{16} GeV$ \cite{pdg}. 
This choice of parameters requires the redefinition $g_{Y}=\sqrt{\frac{3}{5}}g_{1},  g_{B-L}=\sqrt{\frac{3}{4}}g_{4} $.
One then finds, at an arbitrary scale $t=ln(\frac{\mu}{M_{u}})$, that
\begin{equation}
g_{a}(t)^{2}= \frac{g(0)^{2}}{1-\frac{g(0)^{2} b_{a} t}{8 {\pi}^{2}}}, \ a=1,\dots,4 \quad  \ \vec{b}=(\frac{33}{5}, 1, -3, 12) \ .
\label{13}
\end{equation}
These results will be used throughout the analysis. We note that threshold effects and mass splitting between sleptons/squarks will tend to defocus gauge coupling unification. We will ignore these effects in the present paper. Now consider the gaugino masses or, more specifically, the products $g_{a}^{2}|M_{a}|^{2}$, $a=1,\dots,4$ that occur in the beta functions. Denoting the initial values of the gaugino masses by $|M_{a}(0)|$, one finds 
\begin{equation}
g_{a}(t)^{2}|M_{a}(t)|^{2}=\frac{g(0)^{2}|M_{a}(0)|^{2}}{(1-\frac{g(0)^{2} b_{a} t}{8 {\pi}^{2}})^{3}}.
\label{13a}
\end{equation}
Even assuming that the gaugino masses are ``unified'' at $t=0$, making any ratio $\frac{g_{a}(0)^{2}|M_{a}(0)|^{2}}{g_{b}(0)^{2}|M_{b}(0)|^{2}}$ unity, it is clear that the gluino mass contributions will quickly grow to dominate. For example, at the electroweak scale the ratio of the gluino to the $SU(2)_{L}$ gaugino terms is $25.6$. In this paper, so as to simplify the calculation and allow for a quasi-analytic solution, we will {\it not} assume unified gaugino masses, instead taking 
$|M_{1}(0)|^{2}, |M_{2}(0)|^{2}, |M_{4}(0)|^{2} \ll |M_{3}(0)|^{2}$. It follows that in beta functions containing a gluino mass term, the other gaugino terms are sub-dominant everywhere in the scaling regime and can be ignored. Recall that ``non-unified'' gaugino masses easily occur in string vacua, while unification requires additional ``minimal'' criteria \cite{V, 7D}. These are not generically satisfied in our MSSM theory. A similar justification can be made for ignoring soft cubic scalar interactions as sub-dominant.

Next, we make a specific choice for the scalar masses at the unification scale 
$M_{u}$. These are taken to be
 \begin{eqnarray}
 m_{H}(0)^{2} & = &m_{\bar{H}}(0)^{2}, \quad m_{Q_{i}}(0)^{2}=m_{u_{j}}(0)^{2}=m_{d_{k}}(0)^{2}, 
 \nonumber \\
& &  m_{L_{i}}(0)^{2}  = m_{e_{j}}(0)^{2} \neq m_{\nu_{k}}(0)^{2} 
 \label{16}
 \end{eqnarray}
for all $i,j,k=1,2,3$.
Note that the sneutrino masses are different than those of the remaining sleptons. {\it This asymmetry is one ingredient in breaking $U(1)_{B-L}$ at an appropriate scale}. Other than that,
 this choice is taken so as to simplify the RGEs as much as possible and to allow a quasi-analytic solution. We point out that soft scalar masses need not be ``universal'' in string theories, since they are not generically ``minimal''. We emphasize that a $B$-$L$/electroweak hierarchy is possible for a much wider range of initial parameters.
 
Since the $U(1)_{B-L}$ symmetry should  be spontaneously broken by right-handed sneutrinos at energy-momenta larger than the electroweak scale, we begin by restricting the analysis to the slepton sector. This is possible, in part, because initial conditions (\ref{16}) allow a decoupling of sleptons from the squarks and Higgs fields in the RGEs . These fields will be discussed later. 
Subject to the initial conditions (\ref{16}) and associated assumptions, we find that 
\begin{eqnarray}
m_{L_{i}}(t)^{2} & = & m_{L_{i}}(0)^{2}+\frac{1}{6}(1-(1-\frac{g(0)^{2}b_{4}t}{8 {\pi}^{2}})^{-9/4b_{4}})
{\cal{S}'}_{1}(0),  \nonumber \\
m_{e_{i}, \nu_{i}}(t)^{2} & = & m_{e_{i}, \nu_{i}}(0)^{2}-\frac{1}{6}(1-(1-\frac{g(0)^{2}b_{4}t}{8 {\pi}^{2}})^{-9/4b_{4}}) 
{\cal{S}'}_{1}(0) \label{18} 
\end{eqnarray}
where
\begin{equation}
{\cal{S}'}_{1}(0) =  \sum_{i=1}^{3}(-m_{L_{i}}(0)^{2}+m_{\nu_{i}}(0)^{2}) \neq 0 \ .
\label{20}
\end{equation}
Note that in deriving (\ref{18}), we have assumed $|M_{1}(0)|^{2}, |M_{2}(0)|^{2}, |M_{4}(0)|^{2} 
\ll {\cal{S}'}_{1}(0)$. 
Using (\ref{13a}) and (\ref{18}), it follows that the hypercharge, $SU(2)_{L}$ and $B$-$L$ gaugino terms 
are sub-dominant to $g_{4}^{2}{\cal{S}'}_{1}$ at any scale.
Hence, even in the slepton beta functions, which do not have a gluino contribution, the gaugino terms can be ignored..

Given these results, one can now consider $U(1)_{B-L}$ breaking at scales on the order of 
$10^{4}GeV$ or, equivalently, at $t_{B-L} \simeq -28.7$ . We begin by discussing the quadratic mass terms near the origin of field space. The relevant  part of the scalar potential is $V=V_{2s} + \frac{1}{2} D_{B-L}^{2}$,
where $V_{2s}$ and $D_{B-L}$ are given in (\ref{9}) and (\ref{7}) respectively. Recall that the FI term is absorbed into the definition of the soft mass parameters. Expanding this, the slepton quadratic terms at any scale $t$ are 
\begin{equation}
V_{m_{sleptons}^{2}}=\sum_{i=1}^{3} (m_{L_{i}}^{2}|L_{i}|^{2}+m_{e_{i}}^{2}|e_{i}|^{2}+m_{\nu_{i}}^{2}|\nu_{i}|^{2}) \ ,
\label{25a}
\end{equation}
where $i=1,2,3$ and the slepton masses are given by (\ref{18}),(\ref{20}).
The first requirement for spontaneous $B$-$L$ breaking is that at least one of the slepton effective squared masses becomes negative at $t_{B-L}$. Clearly, this cannot happen for $m_{L_{i}}(t_{B-L})^{2}$, which is always positive. However, if the initial squared masses are sufficiently small and 
${\cal{S}'}_{1}(0)$ sufficiently large, both $m_{e_{i}}(t_{B-L})^{2}$ and
$m_{\nu_{i}}(t_{B-L})^{2}$ can become negative. Since the $e_{i}$ fields are electrically charged, we do not want them to get a VEV and, hence, we want  
$m_{e_{i}}(t_{B-L})^{2}$ to be positive. On the other hand, the $\nu_{i}$ fields are neutral in all quantum numbers except $B$-$L$. Hence, if they get a nonzero VEV this will spontaneously break $B$-$L$ at $t_{B-L}$, but leave the $SU(3)_{C} \times SU(2)_{L} \times U(1)_{Y}$ gauge symmetry unbroken. This is indeed possible for a wide range of initial parameters. For simplicity, let us choose the initial right-handed slepton masses to satisfy
\begin{eqnarray}
& &m_{\nu_{1}}(0) =  m_{\nu_{2}}(0)=Cm_{\nu}(0), \quad m_{\nu_{3}}(0)=m_{\nu}(0), \nonumber \\
& & \quad \ \ \ \ m_{e_{1}}(0) = m_{e_{2}}(0)=m_{e_{3}}(0)=Am_{\nu}(0)
\label{27}
\end{eqnarray}
which imply, using (\ref{16}) and (\ref{20}), that  
\begin{equation}
{\cal{S}'}_{1}(0)=(1+2C^{2}-3A^{2})m_{\nu}(0)^{2} \ .
\label{28}
\end{equation}
Taking, for specificity, $A=\sqrt{6}$ and  $C \simeq 9.12$, then  
\begin{equation}
{\cal{S}'}_{1}(0)=149 \ m_{\nu}(0)^{2}
\label {russia1}
\end{equation}
and it follows from (\ref{13}), (\ref{16}), (\ref{18}) and (\ref{27}) that
\begin{eqnarray}
m_{\nu_{1,2}}(t_{B-L})^{2}  \simeq 78.2 \ m_{\nu}(0)^{2}, \quad m_{\nu_{3}}(t_{B-L})^{2} & = & -4m_{\nu}(0)^{2}, \nonumber \\
m_{L_{i}}(t_{B-L})^{2}  =  11 m_{\nu}(0)^{2} , \quad  m_{e_{i}}(t_{B-L})^{2} & = & m_{\nu}(0)^{2}   \label{30} 
\end{eqnarray}
for $1=1,2,3$. We conclude from (\ref{30}) that, near the origin of field space, there are positive quadratic mass terms in the  $L_{i}$, $e_{i}$  and $\nu_{1,2}$ field directions for $i=1,2,3$.  However, 
$m_{\nu_{3}}(t_{B-L})^{2}$ is negative, suggesting a non-zero VEV in the $\nu_{3}$ direction.

To determine this, one must minimize the complete potential $V=V_{2s} + \frac{1}{2} D_{B-L}^{2}$ for the slepton fields. Restricted to these scalars, we find that the vacuum specified by
\begin{equation}
\langle \nu_{1,2} \rangle  =  0, \quad \langle \nu_{3} \rangle= \frac{2m_{\nu}(0)}{\sqrt{\frac{3}{4}}g_{4}},\quad \langle L_{i} \rangle =   \langle e_{i} \rangle = 0
\label{31}
\end{equation}
with $i=1,2,3$ is a local minimum of $V$. The slepton masses at this VEV are
\begin{eqnarray}
& &\langle m_{\nu_{1,2}}^{2} \rangle \simeq 82.2 \ m_{\nu}(0)^{2}, \quad \langle m_{\nu_{3}}^{2} \rangle =  8m_{\nu}(0)^{2},  \nonumber \\
& & \quad  \langle m_{L_{i}}^{2} \rangle  =  7 m_{\nu}(0)^{2}, \quad \langle m_{e_{i}}^{2} \rangle  =  5 m_{\nu}(0)^{2} \ .
\label{32}
\end{eqnarray}
Vacuum (\ref{31}) spontaneously breaks the gauged $B$-$L$ symmetry giving the  $B$-$L$ vector boson a mass, 
\begin{equation}
M_{A_{B-L}}= 2\sqrt{2}m_{\nu}(0) \ ,
\label{33} 
\end{equation}
while preserving the remaining $SU(3)_{C} \times SU(2)_{L} \times U(1)_{Y}$ gauge group. Note that this result is quite robust, and should be applicable to any theory containing at least two right-handed sneutrinos.

We now include the Higgs fields and squarks, and analyze their masses at $t_{B-L}$ around vacuum (\ref{31}). To the order we are working, 
\begin{equation}
m_{\bar{H}}^{2} \simeq m_{H}(0)^{2} \ .
\label{44}
\end{equation}
Using the previous assumptions, the hierarchy of Yukawa couplings, choosing
\begin{equation}
m_{Q_{3}}(0)^{2}=\frac{m_{H}(0)^{2}}{2} \ ,
\label{plane1}
\end{equation}
and requiring that $m_{H}^{2}$ be positive at all relevant scales,  we find that
\begin{equation}
m_{H}^{2} \simeq m_{H}(0)^{2}
e^{  - \frac{3}{4{\pi}^{2}} \int_{t}^{0} {|\lambda_{u_{3}}|^{2}     }      
(1  +[ \frac{-\frac{2}{3 {\pi}^{2}}   
\int_{0}^{t'}{g_{3}^{2}|M_{3}|^{2}}
}{m_{H}^{2}} ] ) } \ .
\label{43}
\end{equation}
Since $\lambda_{u_{3}}$ scales slowly, we take it to be constant with its phenomenological value 
\begin{equation}
\lambda_{u_{3}}(0)=1 \ .
\label{43A}
\end{equation}
As discussed previously, a non-vanishing supersymmetric $\mu$ parameter can arise from non-perturbative effects in the moduli and hidden sector. To simplify the calculations and focus on the $B$-$L$/electroweak hierarchy, we will, henceforth, assume that the $\mu$ parameter, while non-zero, is sufficiently smaller than the electroweak scale so that its effects are sub-dominant. Once this is implemented at one scale, it remains true over the entire scaling regime.
Then, under the previous assumptions, the quadratic pure Higgs potential arises solely from (\ref{9}) and is given by $V_{m_{\rm Higgs}^{2}}= m_{H}^{2} |H|^{2}+ m_{\bar{H}}^{2} |\bar{H}|^{2}-B(H{\bar{H}}+hc)$,
where $m_{H}^{2}$, $m_{\bar{H}}^{2}$  are given in (\ref{43}), (\ref{44}) and $B$ satisfies a relatively simple RGE that won't be discussed here. Henceforth, we assume that for $t \ll 0$ the coefficient $B$ is such that
\begin{equation}
4\left( \frac{B}{m_{\bar{H}}^{2}-m_{H}^{2}}\right)^{2} \ll 1\ .
\label{46}
\end{equation}
This is easily arranged by adjusting $B(0)$. The Higgs mass matrix can then be diagonalized to $V_{m_{\rm Higgs}^{2}}= m_{H'}^{2} |H'|^{2}+ m_{{\bar{H}}'}^{2} |{\bar{H}}'|^{2} $,
where
\begin{equation}
m_{H'}^{2} \simeq m_{H}^{2}-m_{\bar{H}}^{2}\left( \frac{B}{m_{\bar{H}}^{2}-m_{H}^{2}}\right)^{2}, \quad
m_{{\bar{H}}'}^{2} \simeq m_{\bar{H}}^{2}-m_{H}^{2}\left( \frac{B}{m_{\bar{H}}^{2}-m_{H}^{2}}\right)^{2}
\label{48}
\end{equation}
and
\begin{equation}
H' \simeq H - \left( \frac{B}{m_{\bar{H}}^{2}-m_{H}^{2}}\right) {\bar{H}}^*, \quad 
{\bar{H}}' \simeq  \left( \frac{B}{m_{\bar{H}}^{2}-m_{H}^{2}}\right)H^* +{\bar{H}} \ .
\label{49}
\end{equation}
It follows from (\ref{44}), (\ref{46}), and (\ref{48}) that for any $t \ll 0$
\begin{equation}
m_{{\bar{H}}'}^{2} \simeq m_{\bar{H}}^{2}=m_{H}(0)^{2} > 0 \ .
\label{50}
\end{equation}
Importantly, however, we see from (\ref{43}), (\ref{48}) that as $t$ becomes more negative $m_{H}^{2}$ can approach, become equal to and finally become smaller than $m_{\bar{H}}^{2}( B/ (m_{\bar{H}}^{2}-m_{H}^{2}))^{2}$. As discussed shortly, our requirement that $m_{H}^{2}$ be positive forces $m_{H'}^{2}$ to vanish below $t_{B-L}$.  We conclude that at the B-L scale and evaluated at vacuum (\ref{31}), the Higgs masses are
 \begin{equation}
\langle m_{H'}^{2} \rangle > 0, \quad \langle m_{{\bar{H}}'}^{2} \rangle \simeq  m_{H}(0)^{2} 
 \label{56}
 \end{equation}
and, hence, electroweak symmetry is not yet broken.

Now include the squarks and analyze their masses at $t_{B-L}$ around vacuum (\ref{31}). Within the assumptions and approximations discussed earlier, it is straightforward to solve the renormalization group equations for the squarks at arbitrary scale $t$. The simplest are given by
\begin{eqnarray}
m_{Q_{1,2}}^{2} & \simeq & -\frac{2}{3 {\pi}^{2}} \int_{0}^{t}{g_{3}^{2}|M_{3}|^{2}}
+\frac{1}{64 {\pi}^{2}} \int_{0}^{t}{g_{4}^{2}{\cal{S}'}_{1}} +\frac{m_{H}(0)^{2}}{2} \ , \nonumber \\
m_{u_{1,2}, \ d_{i}}^{2} & \simeq & -\frac{2}{3 {\pi}^{2}} \int_{0}^{t}{g_{3}^{2}|M_{3}|^{2}}
-\frac{1}{64 {\pi}^{2}} \int_{0}^{t}{g_{4}^{2}{\cal{S}'}_{1}} +\frac{m_{H}(0)^{2}}{2}    \label{34} 
\end{eqnarray}
where $i=1,2,3$ and 
\begin{eqnarray}
-\frac{2}{3 {\pi}^{2}} \int_{0}^{t}{g_{3}^{2}|M_{3}|^{2}} & = & -\frac{8}{3b_{3}}( \frac{1}{(1-\frac{g(0)^{2}b_{3}t}{8 {\pi}^{2}})^{2}}-1)|M_{3}(0)|^{2} \ , \label{35} \\
-\frac{1}{64 {\pi}^{2}} \int_{0}^{t}{g_{4}^{2}{\cal{S}'}_{1}} & = & -\frac{1}{18}( \frac{1}{(1-\frac{g(0)^{2}b_{4}t}{8 {\pi}^{2}})^{\frac{9}{4b_{4}}}}-1)149 \  m_{\nu}(0)^{2}  \ . \label{36}
\end{eqnarray}
Note that both integrals (\ref{35}) and (\ref{36}) are positive for $t < 0$. Somewhat more complicated are 
\begin{eqnarray}
m_{Q_{3}}^{2} & \simeq & \frac{1}{3}m_{H}^{2}-\frac{2}{3 {\pi}^{2}} \int_{0}^{t}{g_{3}^{2}|M_{3}|^{2}}+\frac{1}{64 {\pi}^{2}} \int_{0}^{t}g_{4}^{2}{\cal{S}'}_{1}+\frac{1}{6}m_{H}(0)^{2} \ , 
  \nonumber \\
m_{u_{3}}^{2} & \simeq & \frac{2}{3}m_{H}^{2}-\frac{2}{3 {\pi}^{2}} \int_{0}^{t}{g_{3}^{2}|M_{3}|^{2}}-\frac{1}{64 {\pi}^{2}} \int_{0}^{t}{g_{4}^{2}{\cal{S}'}_{1}} -\frac{1}{6}m_{H}(0)^{2}  \ . \label{39}
\end{eqnarray}
The masses in (\ref{34}) and (\ref{39}) depend, a priori, on three independent initial parameters, $M_{3}(0)$, $m_{\nu}(0)$ and $m_{H}(0)$. We will relate them as follows. It is clear from (\ref{43}) that 
for $m_{H}^{2}$ to have the appropriate behavior at the electroweak scale fixes $M_{3}(0)$ relative to $m_{H}(0)$.
This will be discussed below. Here, we simply use the result that
\begin{equation}
|M_{3}(0)|^{2}=.0352\ m_{H}(0)^{2} \ .
\label{russia2}
\end{equation}
It is also essential that color and charge be unbroken at the electroweak scale. If we further require that this be the case for all scales $t$, then, as will be discussed shortly, one finds 
\begin{equation}
m_{\nu}(0)^{2}= 0.864 \  m_{H}(0)^{2} \ .
\label{russia3}
\end{equation}
In both (\ref{russia2}) and (\ref{russia3}) we present only the leading, $( B/ (m_{\bar{H}}^{2}-m_{H}^{2}))^{2}$ independent results for these quantities.

For these restricted parameters, the $m_{H}^{2}$ contributions to (\ref{39}) are small and can be ignored. 
Furthermore, at $t_{B-L}$ the Higgs fields have vanishing VEVs. Hence, we can compute the squark masses at (\ref{31}) using the relevant terms in $V=V_{2s} + \frac{1}{2} D_{B-L}^{2}$ . The squark contribution to the quadratic potential is $V_{m_{\rm squark}^{2}}=\langle m_{Q_{i}}^{2} \rangle |Q_{i}|^{2}+\langle m_{u_{i}}^{2} \rangle |u_{i}|^{2}+\langle m_{d_{i}}^{2} \rangle |d_{i}|^{2}$ with
\begin{eqnarray}
\langle m_{Q_{i}}^{2} \rangle & = & m_{Q_{i}}^{2}+\frac{1}{4}g_{4}^{2}|\langle \nu_{3} \rangle |^{2},
\label{41} \\
\langle m_{u_{i},d_{i}}^{2} \rangle &= & m_{u_{i}, d_{i}}^{2}-\frac{1}{4}g_{4}^{2}|\langle \nu_{3} \rangle |^{2}. \nonumber 
\end{eqnarray}
Using (\ref{34}), (\ref{39}) as well as (\ref{13}) and (\ref{31}), we find that at $t_{B-L}$ these squared masses are given by 
\begin{eqnarray}
& & \quad  \langle m_{Q_{1,2}}^{2} \rangle \simeq 0.408 \ m_{\nu}(0)^{2}, \quad   \langle m_{Q_{3}}^{2} \rangle \simeq 0.0435 \ m_{\nu}(0)^{2},  \nonumber \\ 
& &\langle m_{u_{1,2}}^{2}\rangle= \langle m_{d_{i}}^{2} \rangle  \simeq 1.08 \ m_{\nu}(0)^{2}, \quad
 \langle m_{u_{3}}^{2}\rangle \simeq  0.353 \ m_{\nu}(0)^{2}
\label{42}
\end{eqnarray}
for $i=1,2,3$. Note that they are all positive.
It follows from 
(\ref{42}) and (\ref{56}) that (\ref{31}) is indeed a stable, local minimum with respect to all scalar fields at $t_{B-L}$.

Let us now scale down further to the electroweak scale of order $10^{2}GeV$ or, equivalently, $t_{EW} \simeq -33.3$. We simplify the notation and implement (\ref{46}) by taking
\begin{equation}
{\cal{T}}^{2} \equiv \left( \frac{B}{m_{\bar{H}}^{2}-m_{H}^{2}}\right)^{-2} \stackrel{>}{\sim} 40 \ ,
\label{54}
\end{equation}
and choose $M_{3}(0)$ so that
\begin{equation}
m_{H}^{2}=(1-{\Delta}^{2})\frac{m_{H}(0)^{2}}{{\cal{T}}^{2}}, \quad t=t_{EW} 
\label{51}
\end{equation}
for $0 < {\Delta}^{2} < 1$. The upper bound on $\Delta^{2}$ arises from our requirement that $m_{H}^{2}$ be positive for all $t \geq t_{EW}$. Using the previous  assumptions and a numerical solution of (\ref{43}), we find that $m_{H}^{2}$ satisfies condition (\ref{51}) if we choose
\begin{equation}
|M_{3}(0)|^{2}=.0352(1-\frac{11.5(1-{\Delta}^{2})}{{\cal{T}}^{2}})m_{H}(0)^{2} \ .
\label{cat1}
\end{equation}
This justifies (\ref{russia2}) where, for simplicity, we dropped the weak ${\cal{T}}^{2}$ dependence.
It follows from (\ref{48}), (\ref{54}) and (\ref{51}) that at $t_{EW}$
\begin{equation}
m_{H'}^{2} = -{\Delta}^{2}\frac{m_{H}(0)^{2}}{{\cal{T}}^{2}} \ .
\label{59}
\end{equation}
Clearly electroweak breaking can only occur for positive $\Delta^{2}$,  explaining our lower bound on this parameter. Scaling (\ref{59}) up to $t_{B-L}$, we find that the constraint that $\Delta^{2}$ be less than unity implies $m_{H'}^{2}>0$, as claimed in (\ref{56}).

To explore the breaking of electroweak symmetry, one must now consider the complete Higgs potential,
$V=V_{m_{Higgs}^{2}}+\frac{1}{2}D_{Y}^{2}+\frac{1}{2} \sum_{a=1}^{3} D_{SU(2)_{L}a}^{2} $,
at $t=t_{EW}$. 
We express this in terms of the mass diagonal fields $H'$ and ${\bar{H}}'$ defined in (\ref{49}), 
drop quartic terms of ${\cal{O}}({\cal{T}}^{-1})$ and write $H'=(H'^{+}, H'^{0}), {\bar{H}}' =( {\bar{H}}'^{0},  {\bar{H}}'^{-})$.
This potential is easily minimized to give
\begin{eqnarray}
{\langle \langle} H'^{0} \rangle \rangle  =  \frac{2 \Delta \ m_{H}(0)}{{\cal{T}}\sqrt{\frac{3}{5}g_{1}^{2}+g_{2}^{2}}}, \quad \langle \langle H'^{+} \rangle \rangle = \langle \langle {\bar{H}}'^{0} \rangle \rangle  =  \langle \langle {\bar{H}}'^{-} \rangle \rangle = 0 ,\label{65}
\end{eqnarray}
where the double bracket $\langle \langle \ \rangle \rangle$ indicates the vacuum at $t_{EW}$. The Higgs masses evaluated at this VEV are found to be
\begin{equation}
\langle \langle m_{H'^{0}}^{2} \rangle \rangle= 4\frac{{\Delta}^{2} \ m_{H}(0)^{2}}{{\cal{T}}^{2}}, \quad 
\langle \langle m_{{\bar{H}}'^{0}}^{2} \rangle \rangle= \langle \langle m_{{\bar{H}}'^{-}}^{2} \rangle \rangle\simeq m_{H}(0)^{2} \ .
\label{66}
\end{equation}
The three non-radial component fields in $H'$ are the Goldstone bosons associated with the breakdown of electroweak symmetry. They are eaten by the Higgs mechanism to give mass to the $W^{\pm}$ and $Z$ bosons. For example, the $Z$ mass is
\begin{equation}
M_{Z}= \frac{\sqrt{2} \Delta \  m_{H}(0)}{\cal{T}} \simeq 91GeV \ .
\label{67}
\end{equation}

Although the mass eigenstate basis $H'$, ${\bar{H}}'$ is the most natural for analyzing this vacuum, it is of some interest to express it in terms of the original 
$H$ and $\bar{H}$ fields. Using (\ref{49}), we find 
\begin{equation}
\langle \langle H^{+} \rangle \rangle = \langle \langle {\bar{H}}^{-} \rangle \rangle = 0 
\label{5067}
\end{equation}
and, to leading order, that
\begin{equation}
\langle \langle H^{0} \rangle \rangle =\frac{2 \Delta \  m_{H}(0)}{{\cal{T}}\sqrt{\frac{3}{5}g_{1}^{2}+g_{2}^{2}}}, \quad \langle \langle {\bar{H}}^{0} \rangle \rangle = \frac{1}{{\cal{T}}} \langle \langle H^{0} \rangle \rangle \ .
\label{68}
\end{equation}
Note that the condition $\langle \langle {\bar{H}}'^{0} \rangle \rangle =0$ in (\ref{65}) does {\it not} imply the vanishing of $ \langle \langle {\bar{H}}^{0} \rangle \rangle$. Rather, $\langle \langle {\bar{H}}^{0} \rangle \rangle$ is non-zero and related to $\langle \langle H^{0} \rangle \rangle$ through the ratio
 \begin{equation}
 \frac{ \langle \langle H^{0} \rangle \rangle}{\langle \langle {\bar{H}}^{0} \rangle \rangle }\equiv tan\beta
={\cal{T}} +{\cal{O}}({\cal{T}}^{-1}) \ .
\label{69}
\end{equation}
We have indicated the ${\cal{O}}({\cal{T}}^{-1})$  contribution to emphasize that although $tan\beta={\cal{T}}$ {\it to leading order}, this relationship breaks down at higher order in ${\cal{T}}^{-1}$.
We conclude that electroweak symmetry is spontaneously broken at scale $t_{EW}$ by the non-vanishing $H'^{0}$ vacuum expectation value in (\ref{65}). This vacuum has a non-vanishing value of $tan\beta$ which, using the assumption for ${\cal{T}}^{2}$ given in (\ref{54}), satisfies
\begin{equation}
tan\beta \stackrel{>}{\sim} 6.32 \ .
\label{71}
\end{equation}
As far as the Higgs fields are concerned, the vacuum specified in (\ref{65})
is a stable local minimum. As a check on our result, choose $\mu^{2} \sim {\cal{O}}({\cal{T}}^{-4})$ or smaller, that is, non-vanishing but sub-dominant in all equations.  Then (\ref{50}),(\ref{54}) and (\ref{51}) satisfy the constraint equations, given, for example, in \cite{V}, for the Higgs potential to be bounded below and have a negative squared mass at the origin. Furthermore, to the order in ${\cal{T}}^{-1}$ we are working, (\ref{67}) and (\ref{68}) for the Higgs vacuum satisfy the minimization conditions in \cite{V}.

To understand the complete stability of this minimum, it is essential to extend this analysis to the entire field space; that is, to include all sleptons and squarks as well as the Higgs fields. The relevant part of the potential energy is the sum of $V_{2s}$ and the $D_{B-L}$, $D_{Y}$  and $D_{SU(2)_{L}a}$ contributions. The coefficients in this potential are to be evaluated at $t_{EW}$. We find a local extremum at
\begin{eqnarray}
\langle \langle \nu_{1,2} \rangle \rangle & = & 0, \quad \langle \langle \nu_{3} \rangle \rangle= (1.05) \frac{2m_{\nu}(0)}{\sqrt{\frac{3}{4}}g_{4} },\quad \langle \langle L_{i} \rangle \rangle =  \langle \langle e_{i} \rangle \rangle= 0 , \label{72} \\ 
\langle \langle H'^{0} \rangle \rangle & = & \frac{ 2 \Delta \  m_{H}(0)}{tan\beta\sqrt{\frac{3}{5}g_{1}^{2}+g_{2}^{2}}}, \  \langle \langle H'^{+} \rangle \rangle  =  \langle \langle {\bar{H}}'^{0} \rangle \rangle  =  \langle \langle {\bar{H}}'^{-} \rangle \rangle = 0 \nonumber
\end{eqnarray}
for $i=1,2,3$.
To guarantee that this is a stable local minimum, we must compute all of the scalar squared masses at this VEV.  The Higgs masses were given in (\ref{66}). The slepton masses in (\ref{32}) and the squark masses in (\ref{42}) are corrected in two ways, First, they must be scaled down from $t_{B-L}$ to $t_{EW}$. Secondly, they are altered by the non-zero Higgs VEVs. Finally, in addition to (\ref{cat1}) which relates $M_{3}(0)$ to $m_{H}(0)$, one must express $m_{\nu}(0)$ in terms of  $m_{H}(0)$. An overly restrictive but simple way to do this is the following. Demand that,  for any choice of $tan\beta$ and $\Delta$,  all squark and slepton mass squares are non-negative, and, hence, color and electric charge are unbroken, for all values of $t$. We then find that
\begin{equation}
m_{\nu}(0)^{2}=0.864(1- \frac{2.25(1-\Delta^{2})}{{\cal{T}}^{2}} )m_{H}(0)^{2} \ .
\label{russia4}
\end{equation}
This justifies (\ref{russia3}) where, for simplicity,  we dropped the weak ${\cal{T}}^{2}$ dependence. Using the above approximations and dropping appropriate terms of order ${\cal{T}}^{-2}$, the slepton and squark mass squares are given by
\begin{eqnarray}
& & \qquad \ \langle \langle m_{\nu_{1,2}}^{2} \rangle \rangle  \simeq  82.2 \ m_{H}(0)^{2} , \quad
\langle \langle m_{\nu_{3}}^{2} \rangle \rangle \simeq 8.75 \ m_{H}(0)^{2},  \label{74} \\
& &\quad \langle \langle m_{N_{i}}^{2} \rangle \rangle    \simeq  \langle \langle m_{E_{i}}^{2} \rangle \rangle  \simeq  \ 7.00 \ m_{H}(0)^{2}, \ \langle \langle m_{e_{i}}^{2} \rangle \rangle \simeq 5.00 \ m_{H}(0)^{2} \nonumber
\end{eqnarray}
and
\begin{eqnarray}
& & \ \ \qquad  \qquad \langle\langle m_{U_{3}}^{2} \rangle\rangle  \simeq  \langle \langle m_{D_{3}}^{2} \rangle\rangle \simeq 0.132 \ m_{H}(0)^{2}, \nonumber \\
& &\qquad  \qquad \langle\langle m_{U_{1,2}}^{2} \rangle\rangle  \simeq  \langle \langle m_{D_{1,2}}^{2} \rangle \rangle \simeq 0.465 \ m_{H}(0)^{2}, \label{76} \\
& & \langle \langle m_{u_{1,2}}^{2} \rangle \rangle  \simeq  \langle \langle m_{d_{i}}^{2} \rangle \rangle \simeq 1.04 \ m_{H}(0)^{2},  \ 
\langle \langle m_{u_{3}}^{2} \rangle \rangle  \simeq  0.374 \ m_{H}(0)^{2} \nonumber
\end{eqnarray}
for $i=1,2,3$ respectively. Note that the third family up-squark mass squares receive a positive F-term contribution from their Yukawa interaction in (\ref{5}). Although this contribution is of order ${\cal{T}}^{-2}$ and, hence, ignored in (\ref{76}), it can be a sizable correction for smaller values of $tan\beta$.
Since all scalar masses in (\ref{66}), (\ref{74}) and (\ref{76}) are positive, we conclude that the vacuum given in (\ref{72}) is a stable, local minimum of the potential energy.

The vacuum state (\ref{72}) spontaneously breaks both $B$-$L$ and electroweak symmetry, and exhibits a distinct hierarchy between the two. Using (\ref{russia3}), we see that the ratio of the vacuum expectation values is 
\begin{equation}
\frac{\langle \langle \nu_{3} \rangle \rangle}{\langle \langle H'^{0} \rangle \rangle} \simeq (0.976) \frac{\sqrt{\frac{3}{5}g_{1}^{2}+g_{2}^{2}}}{\sqrt{\frac{3}{4}}g_{4}} \frac{tan\beta}{\Delta} \ ,
\label{77}
\end{equation}
where the gauge parameters are computed at $t_{EW}$. Choosing, for specificity, 
$\Delta=\frac{1}{\sqrt{2}}$ and evaluating (\ref{77}) in the region $6.32 \leq tan\beta \leq 40$, we find that
\begin{equation}
19.9 \leq \frac{\langle \langle \nu_{3} \rangle \rangle}{\langle \langle H'^{0} \rangle \rangle} \leq 
126 \ .
\label{78}
\end{equation}

A second measure of the $B$-$L$/electroweak hierarchy is given by the ratio of the $B$-$L$ vector boson mass to the mass of the $Z$ boson. It follows from (\ref{33}), (\ref{russia3}) and (\ref{67}) that
\begin{equation}
\frac{M_{A_{B-L}}}{M_{Z}}\simeq (1.95)\frac{tan\beta}{\Delta} \ .
\label{79}
\end{equation}
Again, using $\Delta=\frac{1}{\sqrt{2}}$ and evaluating this mass ratio in the range $6.32 \leq tan\beta \leq 40$, one finds
\begin{equation}
17.5 \leq \frac{M_{A_{B-L}}}{M_{Z}} \leq 110 \ .
\label{80}
\end{equation}
Note that if we take $\Delta \rightarrow 1$, the upper bound in our approximation, then $\frac{M_{A_{B-L}}}{M_{Z}}$ is essentially $2tan\beta$, whereas if $\Delta \rightarrow 0$ this mass ratio becomes arbitrarily large. For typical values of $\Delta$, 
we conclude that the vacuum (\ref{72}) exhibits a $B$-$L$/electroweak hierarchy of ${\cal{O}}(10)$ to ${\cal{O}}(10^{2})$ in a physically interesting range of $tan\beta$.

Finally, let us review the reasons for the existence and magnitude of the $B$-$L$/electroweak hierarchy. First, initial conditions (\ref{16}),(\ref{27}) give emphasis to the right-handed sneutrinos by {\it not} requiring their masses be degenerate with the $L_{i}$ and $e_{i}$ soft masses. This enables the ${\cal{S}}'_{1}$ parameter (\ref{28}) not only to be non-vanishing but, in addition, to be large enough to dominate all contributions to the RGEs with the exception of the gluino mass terms. This drives $m_{\nu_{3}}^{2}$ negative and initiates $B$-$L$ breaking at scale $m_{\nu}$. Second, $B$ and $M_{3}$ (hence, $m_{H}(t_{EW})$) are chosen to satisfy constraints (\ref{54}) and (\ref{51}) respectively, 
with $0<\Delta^{2}<1$. This insures electroweak breaking for positive $m_{H}^{2}$ at a scale proportional to $\Delta m_{H}(0)/ {\cal{T}}$.  The large value assumed for ${\cal{T}}$ implies that the non-vanishing VEV is largely in the $H^{0}$ direction, allowing one to identify ${\cal{T}}$, to leading order, with $\tan\beta$. Third, equation (\ref{russia4}) insures that squark/slepton squared masses are positive at {\it all} scales. This guarantees that the electroweak breaking is substantially smaller than the $B$-$L$ scale, with the $B$-$L$/electroweak hierarchy proportional to $\tan\beta/\Delta$.  


\section*{Acknowledgments}
B.A.O. would like to thank Nima Arkani-Hamed and Gil Paz for helpful discussions. B.A.O. is grateful to the Institute for Advanced Study and the Ambrose Monell Foundation for support.
M.A. would like to thank the Institute for Advanced Study for its hospitality.
The work of M.A. and B.A.O. is supported in part by the DOE under contract No. DE-AC02-76-ER-03071.
B.A.O. acknowledges partial support from the NSF RTG grant DMS-0636606.

\end{document}